# Warm Molecular Gas in the Envelope and Outflow of IRAS 12496−7650 (DK Cha)

T.A. van Kempen[1], M.R. Hogerheijde[1], E.F. van Dishoeck[1], R.Güsten[2], P. Schilke[2], and L.-Å. Nyman[3]

[1]Leiden Observatory, Leiden University Niels Bohrweg 2, 2333 CA, Leiden, Netherlands
[2]Max Planck Institut für Radioastronomie, Auf dem Hügel 69, D-53121, Bonn, Germany
[3]European Southern Observatory, Casille 19001,Santiago, Chile
e-mail: kempen@strw.leidenuniv.nl

**ABSTRACT**

*Aims.* To obtain insight into the physical structure of the warm gas in the inner envelope of protostars and the interaction with the outflow.
*Methods.* Sub-millimeter observations of $^{12}$CO, $^{13}$CO and/or C$^{18}$O in $J$=3–2, $J$=4–3 and $J$=7–6 were obtained with the APEX Telescope towards IRAS 12496-7650, an intermediate mass young stellar object. The data are compared to ISO-LWS observations of CO $J$=14–13 up to $J$=19–18 lines to test the different proposed origins of the CO lines.
*Results.* The outflow is prominently detected in the 3–2 and 4–3 lines, but not seen at similar velocities in the 7–6 line, constraining the temperature in the high-velocity (> 5 km s$^{-1}$ from line center) gas to less than 50 K, much lower than inferred from the analysis of the ISO-LWS data. In addition, no isothermal gas model can reproduce the emission in both the 7–6 and the higher-$J$ ISO-LWS lines. The 7–6 line probably originates in the inner (<250 AU) region of the envelope at ∼150 K. Detailed radiative transfer calculations suggest that the ISO-LWS lines are excited by a different mechanism, possibly related to the larger-scale outflow. All possible mechanisms on scales smaller than 8″ are excluded. High-resolution continuum as well as high−$J$ $^{12}$CO and isotopic line mapping are needed to better constrain the structure of the warm gas in the inner envelope and the interaction with the outflow.

**Key words.** Star formation - low-mass stars - CO - High excitation lines - APEX - astrochemistry

## 1. Introduction

The early stages of star formation are dominated by circumstellar envelopes and bipolar outflows. Both astrophysical phenomena have been observed over a wide range of proto-stellar masses, including low- and very high-mass young stellar objects (YSOs) (Bachiller & Tafalla 1999; Evans 1999). Outflows are an essential ingredient in the formation of stars as it is theorized that large amounts of angular momentum are transported outwards, necessary for effective accretion of the mass stored in the circumstellar envelope onto the central star and circumstellar disk. Resolved circumstellar envelopes have been observed and modelled around both low-mass (e.g., Shirley et al. 2000; Jørgensen et al. 2002) and high-mass protostars (e.g., van der Tak et al. 2000; Mueller et al. 2002). The region where the central part of the envelope and the outflow interact is generally not resolved in single-dish observations, but is a crucial link in understanding star formation. High-frequency single-dish observations could probe these regions effectively. This letter presents the first observations of CO 4–3 and 7–6 toward a southern YSO, IRAS 12496-7650 (DK Cha), using APEX.[1]

CO and its isotopologues $^{13}$CO, C$^{18}$O and C$^{17}$O are often used the characterize the envelopes in these early stages (e.g. Jørgensen et al. 2002). The low excitation $J$=3–2, 2–1 and 1–0 lines of the CO isotopologues primarily trace the very cold material in the outer envelope. Highly excited lines of CO in the far-infrared, such as $J$=14–13 to $J$=17–16, probe the warm gas and have been detected with the *Long Wavelength Spectrometer* (LWS) on board the *Infrared Space Observatory* (ISO) toward YSOs (e.g., Giannini et al. 1999; Maret et al. 2002). Two interpretations have been put forward to explain this emission: warm (∼ 200 K) gas in the innermost envelope (e.g., Ceccarelli et al. 1999; Maret et al. 2002) or hot (∼ 750 K) gas in an outflow (e.g., Giannini et al. 2001; Nisini et al. 2002). Because the ISO-LWS data lack both spatial (90″ beam) and spectral

---

[1] This paper is based on data acquired with the *Atacama Pathfinder EXperiment*, which is a collaboration between the Max-Planck-Institüt für Radioastronomie, the European Southern Observatory, and the Onsala Space Observatory

($\lambda/\Delta\lambda \approx 100$) resolution, the origin of this high-$J$ emission could not be firmly identified.

Much higher spatial and spectral resolution is available for intermediate-$J$ CO lines such as $J=6$–5 up to $J=8$–7. These lines are well suited to probe gas of 100–200 K expected in the inner envelopes of protostars and distinguish between these two models. Due to technological challenges and the high opacity of the atmosphere, they are not commonly observed. Hogerheijde et al. (1998, 1999) found surprisingly bright lines of $^{12}$CO 4–3 and 6–5 and $^{13}$CO 6–5 for a number of low-mass YSOs using the *Caltech Submillimeter Observatory* (CSO) in beams of 10–15″, which they interpreted with both outflows ($^{12}$CO) and (photon-)heated envelopes ($^{13}$CO) (Spaans et al. 1995). Since submillimeter continuum observations were still lacking at the time, the data were never interpreted in the context of envelope models.

APEX opens up the possibility to routinely obtain intermediate-$J$ CO lines at high sensitivity in the Southern sky. We present initial APEX data for IRAS 12496-7650. This Class I source is part of the Chamaeleon II star-forming region (∼250 pc) and is its most luminous and massive member, $L_{bol}$=50 L$_\odot$. It has been classified as a Herbig Ae star with a circumstellar envelope and is associated with a weak bipolar outflow (Knee 1992). Continuum studies at infrared and millimeter wavelengths by Henning et al. (1998) give a H$_2$ column density of $1.0\times10^{23}$ cm$^{-2}$ and a gas mass of 0.16 M$_\odot$ from a 1.3 mm flux of 0.5 Jy within a beam of 23″. Spectroscopically, this source was detected by Giannini et al. (1999) with ISO-LWS in CO lines between 14–13 up to 19–18. No other molecular lines, such as H$_2$O or OH, could be identified, but the origin of this hot CO could not be determined. Using the detected lines and a LVG model, two possible sets of physical conditions were proposed: $T$ = 200 or 750 K with a H$_2$ gas density of $4\times10^6$ or $5\times10^5$ cm$^{-3}$, both within a region with a 300 AU radius.

## 2. Observations

The observations were performed in July 2005 during the science verification of the APEX telescope in Chajnantor, Chile, using the APEX-2a receiver for the $^{12}$CO 3–2 (345.795 GHz) line and the FLASH receiver for the higher excited $^{12}$CO lines: 4–3 (461.041 GHz) and 7–6 (806.651 GHz), observed simultaniously. Additional observations were taken of $^{13}$CO and C$^{18}$O, both in 3–2. The APEX telescope, with its 12 m dish, has beam sizes of 18″ at 345 GHz, 13.5″ at 460 GHz and 8″ at 806 GHz. Preliminary beam efficiencies, obtained by the APEX team from observations of planets, are 0.7 (345 GHz), 0.65 (460 GHz) and 0.47 (806 GHz) (Güsten et al., this volume). The Fast Fourier Transform Spectrometer (FFTS) was used as a back-end with a bandwidth of 500 MHz and a velocity resolution of 0.05 (345 GHz) to 0.09 (806 GHz) km s$^{-1}$. Integration times ranged from 5 minutes per line for $^{12}$CO 3–2 to 10 minutes for the FLASH instrument. Typical system temperatures on the sky were ∼1000 K at 460 and ∼4000 K at 810 GHz. Beam switching was used with a switch of 180″. Calibration uncertainties are estimated to be 30%. The coordinates of IRAS 12496-7650 were taken from the Spitzer IRAC image and are RA = 12$^h$53$^m$17$^s$.2, Dec = -77°07′10″.6 (J2000)

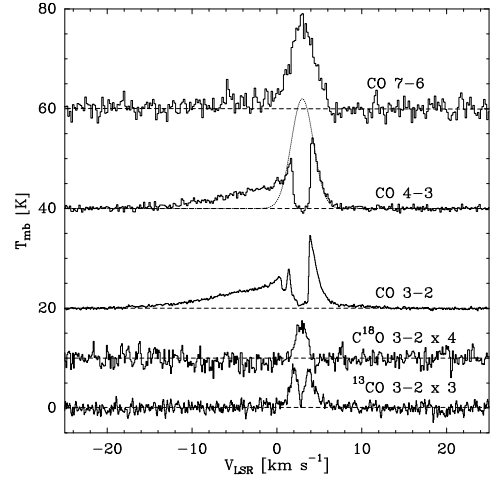

**Fig. 1.** CO observations of IRAS 12496-7650. Spectra have been shifted by +10,+20, +40 and +60 K for C$^{18}$O 3–2,$^{13}$CO 3–2, CO 4–3 and CO 7–6 respectively. C$^{18}$O 3–2,$^{13}$CO 3–2 have been multiplied with a factor of 3 and 4 respectively. The grey line in the 4-3 spectrum is the Gaussian fit to the central component corrected for absorption.

**Table 1.** Observed line strengths

| Line | | $\int T_{MB}V^a$ (K km s$^{-1}$) | Gaussfit (K km s$^{-1}$) | Peak (K) | $\int T_{MB}dV^b$ (K km s$^{-1}$) |
|---|---|---|---|---|---|
| C$^{18}$O | 3–2 | 9.2 | 9.2 | 1.9 | - |
| $^{13}$CO | 3–2 | 6.7 | 12.5 | 3 | - |
| $^{12}$CO | 3–2 | 92.8 | 132 | 25 | 52.6 |
|  | 4–3 | 90.0 | 144 | 23 | 48.6 |
|  | 7–6 | 43.7 | 40.2 | 19.2 | 9.8$^c$ |

$^a$ Total Emission. Integrated between -25 to 20 km s$^{-1}$
$^b$ Outflow emission. Integrated between -25 to 0 km s$^{-1}$
$^c$ Upper limit as some envelope material may be included

Spectra taken at different nights differed by 0.5 km s$^{-1}$ in velocity, making the actual LSR velocity hard to determine. All spectra were carefully aligned fitting Gaussians (see Fig. 1), with the C$^{18}$O 3–2 line at $V_{LSR}$=3.2 km s$^{-1}$ taken to be the source velocity.

## 3. Results and Analysis

The resulting spectra are presented in Fig. 1, and line parameters in Table 1. The CO 4–3, 3–2 and $^{13}$CO 3–2 lines suffer from absorption, probably caused by a combination of cold foreground cloud material, self-absorption from the outer cold envelope and emission of these lines in the off position. Because no absorption is seen in the CO 7–6 or C$^{18}$O 3–2 line profiles, the off-position likely consists of cold, low density gas, similar to any fore-ground material. To estimate the intrinsic source spectrum, Gaussian profiles were fitted to the lines using only the red side of the profile. Results can be seen in column 4 of Table 1. The dividing line between outflow and envelope at <0 km s$^{-1}$ is based on the C$^{18}$O and $^{13}$CO lines, which show only low-velocity emission.

## 3.1. Outflow emission

The CO 3–2 and 4–3 spectra in Fig. 1 have an obvious outflow signature. Remarkably, the CO 7–6 line only shows a very weak sign of outflow material. Either the outflow is not emitting in this transition or its spatial extent is larger than that of the 7–6 beam (8″) and smaller than that of the 4–3 beam (15″). Assuming the (undetected) red wing to have a similar velocity extent, the outflow spans a total FWHM of ∼25 km s$^{-1}$ around the central velocity of ∼3 km s$^{-1}$ in the LSR coordinate frame. It is excluded that the gas is highly optically thick, since no $^{13}$CO 3–2 wing emission is detected at a level of 0.2 K, compared to levels of ∼3 K seen in $^{12}$CO 3–2. This limits the optical depth to ∼2.

The $^{12}$CO line intensity ratios can be used to constrain the outflow parameters. The line ratio as a function of velocity is constant for 4–3/3–2 in a region up to −15 km s$^{-1}$ from the line center and for 7–6/3–2 in a region up to −8 km s$^{-1}$. At higher velocities, noise becomes dominant. The ratios of 1.2 for 4–3/3–2 and ∼0.25 for 7–6/3–2 constrain the temperature to ∼50 K if the gas is thermalized. If the gas is subthermally excited ($n < n_{cr} = 5 \times 10^4$ cm$^{-3}$), higher temperatures are possible up to 200 K for $n \sim 10^4$ cm$^{-3}$. The observed line ratio excludes $n < 10^4$ cm$^{-3}$ for any temperature (following plots by Jansen et al. 1994). An important conclusion is that either the temperature and/or density of the outflowing gas in CO 3–2 and 4–3 are much lower than those needed to explain the CO 14–13 to 19–18 lines (Giannini et al. 1999). Combined with the lack of a prominent wing in CO 7–6, this indicates that the high-velocity gas seen with APEX in $^{12}$CO 4–3 and 3–2 is unlikely to be the same as the warm gas detected by ISO-LWS.

## 3.2. Envelope

The C$^{18}$O intensity can be used to estimate a column density of the material in the envelope. The inferred C$^{18}$O column density is ∼ $5 \times 10^{15}$ cm$^{-2}$, corresponding to a H$_2$ column density of $3 \times 10^{22}$ cm$^{-2}$ in the 18″ beam, assuming LTE, a constant CO/H$_2$ ratio of $10^{-4}$ and a CO/C$^{18}$O isotopologue ratio of 600. This column density is a factor 3 lower than that of Henning et al. (1998), possibly due to dust properties or freeze-out.

The CO 7–6 line does not show a clear outflow signature between $V_{LSR}$=-20 to 0 km s$^{-1}$ and over 80% of its intensity originates at velocities nominally associated with the envelope. Thus, it is natural to investigate whether such a bright high-excitation line can originate in the inner regions of the envelope. Considering that the $J$=7 energy level of CO lies at $T \sim$ 155 K and assuming $T_{kin}$=155K, an optical depth > 3, LTE and spherical symmetry, the emitting region would have a radius of about 325 AU.

To investigate whether the hot gas traced through the high-$J$ CO lines, observed with ISO-LWS, can explain the 7–6 emission, we perform a statistical equilibrium calculation using the RADEX code with updated rate coefficients from Flower (2001) (Schöier et al. 2005). This analysis is similar to that of Giannini et al. (1999) where a warm (∼200 K) or hot (∼750 K) region with a radius of ∼300 AU is needed to reproduce the observed ISO-LWS fluxes. It is assumed that the 3–2 and 4–3 emission traces the cold outer envelope only (see below). A line width of 3.5 km s$^{-1}$ was chosen in the formulation of the escape probability, corresponding to that observed for the 7–6 line. The ortho/para ratio of the collision partner, H$_2$, was taken to be thermalized at the gas temperature. Figure 2 shows the observed APEX and ISO-LWS fluxes together with such a model for $T$=200 K and $n$=4 × 10$^6$ cm$^{-3}$ within a region with a radius of 300 AU. It is clear that this fit gives a flux in the 7–6 line that is a factor of at least 4 too high compared with the observed value. No single temperature-density fit could be found that reproduces both the 7–6 and the ISO-LWS lines, including the hottest model proposed by Giannini et al. (1999) of 750 K. Such a model would still overestimate the 7–6 emission by a factor of 3. This shows that the central region of the envelope is not filled with large quantities of hot gas, otherwise the observed emission in 7–6 would be higher. Even assuming that the origin of the 7–6 line is limited to a central hot region, only 25 % of the ISO-LWS flux would originate in the inner 8″, while all other emission would be located within the 90″ ISO beam, but outside the 8″ beam of APEX.

If such hot gas cannot explain the intensity of the 7–6 line, can the inner warm region of the envelope around the 50 L$_\odot$ YSO reproduce this emisson? And how much can this warm gas contribute to the high-$J$ CO lines? To investigate these questions we calculate the temperature structure in a spherical envelope using the radiative transfer code DUSTY (Ivezić & Elitzur 1997), with relevant parameters as seen in Table 2, based upon previous dust observations (Henning et al. 1998). The modeling strategy by Jørgensen et al. (2002) and Schöier et al. (2002) for other Class I sources is followed and then adapted for the higher luminosity. A drop abundance profile was adopted in which the CO abundance is normal at $10^{-4}$ in the outermost part of the envelope ($n < 10^5$ cm$^{-3}$) and in the inner part ($T > 30$ K), but is orders of magnitude lower in the intermediate cold zone (Jørgensen 2004). A power-law envelope with a slope $\alpha$=1.2 was found to best explain the high 3–2 and 4–3 fluxes relative to the 7–6 flux. These calculations resulted in a temperature profile which is significantly colder than the mean temperature used in the RADEX analysis. At 350 AU, temperatures have already dropped as low as 75 K. The region with temperatures above 200 K has a radius of 40 AU, severely limiting the existence of a hotter inner region capable of producing strong ISO-LWS fluxes. The radiative transfer code RATRAN (Hogerheijde & van der Tak 2000) has subsequently been used to model all CO rotational lines between $J_{up}$=1 and 35. The best fit model is included in Fig. 2 with all CO fluxes with $J_{up} > 7$ calculated for a 8″ beam. Fluxes for lines with $J_{up} \leq 7$ or lower are convolved with the appropriate APEX beam. It is clear that such an envelope model can account for the 7–6, 4–3 and 3–2 emission, but cannot do the same for the observed ISO-LWS fluxes. The observed C$^{18}$O 3–2 is reproduced within 30%. We conclude that the material traced in the 90″ ISO-LWS beam must be distributed over scales >8″. This main conclusion is not changed by a more complex source structure, including a possible clumpy distribution of the gas and dust on scales smaller than 8″ (Giannini et al. 1999).

**Table 2.** Envelope model parameters

| Model parameters | | | |
|---|---|---|---|
| General | | Gas | |
| $L_{bol}$ | 50 L$_\odot$ | $n_{1000AU}$ | $7 \times 10^6$ cm$^{-3}$ |
| $\alpha$ | 1.2 | $X(^{12}CO)$ | $10^{-4}$ for $T > 30$ K |
| $T_{in}$ | 250 K | | or $R >$ 9,000 AU |
| $R_{in}$ | 39 AU | $X(^{12}CO)$ | $10^{-8}$ for $T < 30$ K |
| $R_{out}$ | ∼40,000 AU | | and $R <$ 9,000 AU |

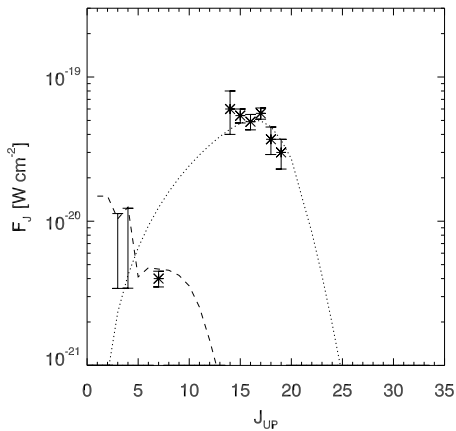

**Fig. 2.** Observed fluxes in various CO lines (points) and the best model fit ($T_{kin}$ = 200 K, n(H$_2$)=4×10$^6$ cm$^{-3}$) of the ISO-LWS data following Giannini et al. (1999) using RADEX (dotted line). Our best fit model using a full radiative transfer model for the envelope with parameters in Table 2 is indicated with a dashed line. Note that the 3–2 and 4–3 lines are indicated by upper and lower limits only. Lower limit is defined by observed emission between 0 and 10 km s$^{-1}$. Upper limit is defined using the correction discussed (see Table 1).

## 4. Conclusions

IRAS 12496-7650, a Herbig Ae star, is accompanied by both a large envelope as traced by the C$^{18}$O $J$=3–2, CO 3–2, 4–3 and CO $J$=7–6 lines and an outflow seen in CO $J$=3–2 and $J$=4–3 wings, reaching speeds up to ±25 km s$^{-1}$. The ratios of the outflow wing emission constrain the gas in the outflow to a kinetic temperature of ∼50 K, or a density of ∼few×10$^4$ cm$^{-3}$, depending on whether the outflow gas is thermalized or not. A spherically symmetric envelope model is able to reproduce the observed 7–6 line emission using a full radiative transfer code. This envelope model extends inwards towards radii of 30 AU and temperatures of 250 K. The ISO-LWS lines do not seem to originate in these physical components and only one possible scenario remains. The emission is produced on scales larger than 8″, but within 90″. The low-excitation lines of CO are dominated by the cold outer envelope.

Future observations such as can be performed at APEX are essential to further probe the inner warm regions of YSOs like IRAS 12496–7650. Both continuum mapping at different sub-mm wavelengths as well as spectral mapping in high-$J$ CO lines (including isotopologues) with instruments like FLASH and CHAMP+ will be able to further distinguish envelope and outflow emission, as well as the energetics of their interaction.

*Acknowledgements.* The authors are grateful to the APEX staff for carrying out the observations. Antonio Crapsi and Jes Jørgensen are thanked for useful discussions. TvK and astrochemistry at Leiden Observatory are supported by a Spinoza prize and grant 614.041.004 from the Netherlands Organization of Scientific Research (NWO).